\documentclass[12pt]{article}

\usepackage{amssymb}
\usepackage{amsmath}
\usepackage{amscd}
\usepackage{latexsym}

\newcommand{\be}{\begin{equation}}
\newcommand{\ee}{\end{equation}}
\newcommand{\Dlt}{\Delta}
\newcommand{\dlt}{\delta}
\newcommand{\prt}{\partial}
\newcommand{\br}{{\bf r}}
\newcommand{\bk}{{\bf k}}
\newcommand{\bt}{\beta}
\newcommand{\vp}{\varphi}
\newcommand{\ep}{\varepsilon}
\newcommand{\al}{\alpha}
\newcommand{\ra}{\rightarrow}
\newcommand{\sgm}{\sigma}

\newcommand{\om}{\omega}
\newcommand{\Om}{\Omega}

\newcommand{\dgr}{\dagger}
\newcommand{\lbd}{\lambda}
\newcommand{\Lbd}{\Lambda}

\newcommand{\lgl}{\langle}
\newcommand{\rgl}{\rangle}

\begin{document}

\begin{center}
{\Large{\bf Particle fluctuations in nonuniform and trapped 
Bose gases} \\ [10mm]
 
V.I. Yukalov} \\ [5mm]

{\it Bogolubov Laboratory of Theoretical Physics, \\
  Joint Institute for Nuclear Research, Dubna 141980, Russia}
\end{center}

\vskip 3cm

\begin{abstract}

The problem of particle fluctuations in arbitrary nonuniform 
systems with Bose-Einstein condensate is considered. This 
includes the case of trapped Bose atoms. It is shown that the 
correct description of particle fluctuations for any nonuniform 
system of interacting atoms always results in thermodynamically 
normal fluctuations.

\end{abstract}

\vskip 2cm

{\parindent=0pt
{\bf Key words}: particle fluctuations; Bose-Einstein condensation; 
nonuniform Bose systems; thermodynamic limit; thermodynamic stability

\vskip 2cm

{\bf PACS}: 03.75.Hh, 03.75.Nt, 05.30.Ch, 05.30.Jp, 67.10.Ba, 
67.10.Fj, 67.85.Bc, 67.85.De, 67.85.Jk
}

\newpage

\section{Trapped Bose atoms}

Trapped gases of Bose atoms constitute an important class of 
nonuniform systems. Thermodynamics and dynamics of trapped atoms 
have  been studied in many details. For review, see the book [1] 
and review articles [2-5]. One of the topics that have met a 
great deal of controversy is the problem of particle fluctuations 
in the systems  with Bose-Einstein condensate. There have appeared 
a number of publications claiming the occurrence of thermodynamically 
anomalous condensate fluctuations everywhere below the condensation 
temperature, for both uniform as well as trapped systems. 
Thermodynamically anomalous fluctuations correspond to the particle 
dispersion proportional to $N^{4/3}$, instead of $N$ for the normal 
fluctuations. As has been explained in the review papers [3,5], the 
occurrence of these thermodynamically anomalous fluctuations implies 
thermodynamic instability. Hence, the systems with such anomalous 
fluctuations simply cannot exist. For the case of uniform systems, 
it has been shown [6-8] that the thermodynamically anomalous 
fluctuations are due to incorrect calculations, while the correct 
calculational procedure yields thermodynamically normal particle 
fluctuations. 

In the present paper, this result is generalized to the case of 
arbitrary nonuniform systems, including the case of trapped atoms. 
It will be shown that for any Bose-condensed system, whether 
uniform or nonuniform, particle fluctuations are always 
thermodynamically normal. The notion of thermodynamically normal 
fluctuations is related to that of thermodynamic limit. For a 
uniform system of $N$ atoms in volume $V$, the thermodynamic limit 
is commonly defined as
\be
\label{1}
N \; \ra \; \infty \; , \qquad V \; \ra \; \infty \; , 
\qquad \frac{N}{V} \; \ra \; const \; .
\ee            
For an arbitrary nonuniform system, the thermodynamic limit can be 
defined [9] as the limit
\be
\label{2}
N \; \ra \; \infty \; , \qquad \langle \hat A \rangle 
\; \ra \; \infty \; , \qquad 
\frac{\langle \hat A \rangle }{N} \; \ra \; const \; ,
\ee
valid for the statistical average $<\hat{A}>$ of any extensive 
observable $\hat{A}$.

Atoms are often trapped by means of a power-law confining potential
\be
\label{3}  
U(\br) = \sum_{\al=1}^d \; \frac{\om_\al}{2} \; \left |
\frac{r_\al}{l_\al} \right |^{n_\al} \; ,
\ee
where $d$ is space dimensionality, $n_\alpha>0$, and the potential 
parameters are related by the equations
\be
\label{4}  
\om_\al = \frac{1}{ml^2_\al} \; , \qquad
l_\al = \frac{1}{\sqrt{m\om_\al} } \;   .
\ee
It is convenient to introduce the {\it effective frequency} and 
the {\it effective localization length}, respectively,
\be
\label{5}  
\om_0 \equiv \left ( \prod_{\al=1}^d \om_\al \right )^{1/d} \; ,
\qquad l_0 \equiv \left ( \prod_{\al=1}^d l_\al \right )^{1/d} \; ,
\ee
which are connected by the relations
$$
\om_0 = \frac{1}{ml_0^2} \; , \qquad 
l_0 = \frac{1}{\sqrt{ml\om_0} } \; .
$$

For this type of the confining potentials, the thermodynamic 
limit (2) reduces [9] to the limit
\be
\label{6}  
N \; \ra \; \infty \; , \qquad \om_0 \; \ra \; 0 \; , 
\qquad N\om_0^s \; \ra \; const   \; ,
\ee
or, equivalently, to
\be
\label{7}  
N \; \ra \; \infty \; , \qquad l_0 \; \ra \; \infty \; , 
\qquad \frac{N}{l_0^{2s}} \; \ra \; const \; ,
\ee 
where the notation of the {\it confining strength}
\be
\label{8}  
s \equiv \frac{d}{2} + \sum_{\al=1}^d \; \frac{1}{n_\al} 
\ee
is introduced.

The passage to the uniform system corresponds to 
$n_\alpha \rightarrow \infty$, when $s \rightarrow d/2$ and 
$l_0 \rightarrow L/2$, with $L$ being the system length, such 
that $L^d = V$. 

The comparison of Eqs. (1) and (7) tells us that, for a 
confined system, the effective volume can be defined [1,9] as 
$V\equiv\; const\cdot l_0^{2s}$. This definition seems to be 
not unique, since the proportionality constant here is not yet 
prescribed. However, we can remember that real trapped systems 
are always finite, being bounded by the trap volume $V$. The 
power-law confining potential (3) is just a model for the real 
trapping potential. For sufficiently large traps, this is a good 
model, which does not contradict the fact that the real trap has 
a finite volume $V$. Therefore, the relation $V \equiv const\cdot 
l_0^{2s}$ can be treated as defining the proportionality 
coefficient.

If $H$ is the system grand Hamiltonian, then the grand thermodynamic  
potential is
\be
\label{9}  
\Om = - P V = - T \ln\; {\rm Tr}\; e^{-\bt H} \; ,
\ee
where $T$ is temperature and $\beta \equiv 1/T$. This defines 
the system pressure
\be
\label{10}  
P = \frac{T}{N} \; \rho \; \ln \; {\rm Tr} \; e^{-\bt H} \; ,
\ee     
in which $\rho \equiv N/V$ is the average density of atoms. Thus, 
remembering that any trap has a finite volume $V$ allows us to use, 
for trapped atoms, the same thermodynamic relations as for uniform 
systems. 

Particle fluctuations are characterized by the dispersion 
\be
\label{11}
\Dlt^2(\hat N) \equiv \langle \hat N^2 \rangle - 
\langle \hat N \rangle^2
\ee
of the number-of-particle operator$\hat{N}$. This dispersion is 
straightforwardly connected with the isothermal compressibility
\be
\label{12}
\kappa_T \equiv - \frac{1}{V} \left (
\frac{\prt P}{\prt V} \right )_T^{-1} =
\frac{\Dlt^2(\hat N)}{\rho TN } \;  ,
\ee 
hydrodynamic sound velocity
\be
\label{13}
s_T \equiv \frac{1}{m} \left ( \frac{\prt P}{\prt\rho} 
\right ) = \frac{1}{m\rho\kappa_T} = 
\frac{NT}{m\Dlt^2(\hat N)} \; ,
\ee
where $m$ is atomic mass, and with the central structure factor
\be
\label{14}
S(0) = T\rho \kappa_T = \frac{T}{ms_T^2} = 
\frac{\Dlt^2(\hat N)}{N} \;  .
\ee   
Equations (12), (13), and (14) are exact, being valid for any 
nonuniform system. In order that these measurable quantities 
would have sense, it is necessary and sufficient that the particle 
fluctuations be thermodynamically normal, such that
\be
\label{15}
\frac{\Dlt^2(\hat N)}{N} \; \ra \; const \qquad (N\ra\infty) \; .
\ee 
If the particle fluctuations would be thermodynamically anomalous, 
being proportional to $N^{4/3}$, as is claimed by some authors, 
the compressibility (12) would be divergent, sound velocity (13) 
would be zero, while the structure factor (14) would be infinite. 
Such a thermodynamically anomalous behavior would mean that the 
considered system is thermodynamically unstable.

\section{Nonuniform Bose systems}

An arbitrary nonuniform Bose system is characterized by the 
energy Hamiltonian
$$
\hat H = \int \hat\psi^\dgr (\br) \left ( - \;
\frac{\nabla^2}{2m} + U \right ) \hat\psi(\br) \; d\br \; + 
$$
\be
\label{16}
+\; \frac{1}{2} \int  \hat\psi^\dgr (\br) \hat\psi^\dgr (\br')
\Phi(\br-\br') \hat\psi(\br') \hat\psi(\br) \; d\br d \br ' \;,
\ee 
in which $\hat{\psi}$ is the Bose field operator, $U=U({\bf r})$ 
is an external potential, such as the trapping potential, and 
$\Phi({\bf r})$ is an integrable symmetric interaction potential. 
Here and in what follows, the units are used where $\hbar=1,\; 
k_B =1$.

The appearance of the Bose-Einstein condensate is the necessary 
and sufficient condition for the global gauge symmetry breaking 
[10]. The latter is conveniently realized by means of the 
Bogolubov shift [11, 12] for the field operator
\be
\label{17}
\hat\psi(\br) = \eta(\br) + \psi_1(\br) \;  .
\ee
Here $\eta ({\bf r})$ is the condensate wave function and 
$\psi_1({\bf r})$ is the Bose field operator of uncondensed 
atoms. To exclude the double counting, these variables are to 
be orthogonal to each other:
\be
\label{18}
\int \eta^*(\br) \psi_1(\br) \; d\br =  0 \; .
\ee

The condensate wave function is normalized to the number of 
condensed atoms
\be
\label{19}
N_0 = \int \rho_0(\br) \; d\br \; , \qquad
\rho_0(\br) \equiv |\eta(\br)|^2 \; .
\ee
And the number of uncondensed atoms is given by the average
\be
\label{20}
N_1 = \lgl \hat N_1 \rgl \; , \qquad  
\hat N_1 \equiv \int \psi_1^\dgr(\br) \psi_1(\br) \; 
d\br \; .
\ee
The total number of atoms in the system is
\be
\label{21}
N = N_0 + N_1 = \lgl \hat N \rgl \; , \qquad 
\hat N = N_0 + \hat N_1 \; .
\ee

The condensate wave function plays the role of the order 
parameter characterizing the gauge symmetry breaking:
\be
\label{22}
\eta(\br) = \lgl \hat\psi(\br) \rgl \; , \qquad
\lgl \psi_1(\br) \rgl = 0 \;  .
\ee
This means that the Hamiltonian should not contain the terms 
linear  in the field operators of uncondensed atoms [13]. 
The latter condition can be realized by complimenting the 
Hamiltonian with a counterterm
\be
\label{23}
\hat\Lbd = \int \left [
\lbd(\br) \psi_1^\dgr(\br) + \lbd^*(\br) \psi_1(\br) \right ]
\; d\br \; ,
\ee
for which
\be
\label{24}
\lgl \hat\Lbd \rgl = 0 \; ,
\ee
and the Lagrange multipliers $\lambda({\bf r})$ are chosen so 
that to kill the terms linear in $\psi_1({\bf r})$.   

Taking into account the statistical constraints (19), (20), and 
(24) defines the grand Hamiltonian
\be
\label{25}
H [\eta,\; \psi_1 ] = \hat H - \mu_0 N_0 -\mu_1 \hat N_1 
- \hat\Lbd \;  ,
\ee 
which is a functional of the field variables $\eta$ and $\psi_1$. 
The quantities $\mu_0$ and $\mu_1$ are the Lagrange multipliers 
guaranteeing the validity of the normalization conditions (19) 
and (20). With the Bogolubov shift (17), Hamiltonian (25) is the 
sum
\be
\label{26}
H [\eta,\; \psi_1 ] = \sum_{n=0}^4 H^{(n)} 
\ee
of the terms labelled according to the entering powers of the 
operators $\psi_1$. The zero-order term
$$
H^{(0)} = \int \eta^*(\br) \left ( - \; \frac{\nabla^2}{2m} +
U - \mu_0 \right ) \eta(\br) \; d\br \; +
$$
\be
\label{27}
+ \; \frac{1}{2} \int \Phi(\br-\br')\; |\eta(\br)|^2 \;
|\eta(\br')|^2 \; d\br d\br '
\ee
does not contain the field operators of uncondensed atoms. The 
first-order term $H^{(1)} = 0$, being eliminated by the linear 
killer (23). The second-order term is
$$
H^{(2)} = \int \psi_1^\dgr(\br) \left ( - \; \frac{\nabla^2}{2m} +
U - \mu_1 \right ) \psi_1(\br) \; d\br \; +
$$
$$
+ \int \Phi(\br-\br') \left [ |\eta(\br)|^2 \psi_1^\dgr(\br')
\psi_1(\br') + \eta^*(\br) \eta(\br') \psi_1^\dgr(\br') \psi_1(\br) +
\right.
$$
\be
\label{28}
\left.
+ \frac{1}{2} \; \eta^*(\br) \eta^*(\br') \psi_1(\br') 
\psi_1(\br) + \frac{1}{2} \; \eta(\br) \eta(\br') 
\psi_1^\dgr(\br') \psi_1^\dgr(\br) \right ] \; d\br d\br ' \; .
\ee
Respectively, the third-order term is
\be
\label{29}
H^{(3)} = \int \Phi(\br-\br') \left [ \eta^*(\br)
\psi_1^\dgr(\br') \psi_1(\br') \psi_1(\br) + 
\psi_1^\dgr(\br) \psi_1^\dgr(\br') \psi_1(\br') \eta(\br)
\right ] \; d\br d\br ' \; ,
\ee
and for the fourth-order term, one has
\be
\label{30}
H^{(4)} =  \frac{1}{2} \int \psi_1^\dgr(\br) \psi_1^\dgr(\br') 
\Phi(\br-\br') \psi_1(\br') \psi_1(\br) \; d\br d\br ' \; .
\ee

The evolution equations for the variables $\eta$ and $\psi_1$ can 
be represented by the variational forms
\be
\label{31}
i \; \frac{\prt}{\prt t} \; \eta(\br,t) = 
\lgl \frac{\dlt H[\eta,\psi_1]}{\dlt\eta^*(\br,t)} \rgl
\ee
and 
\be
\label{32}
i \; \frac{\prt}{\prt t} \; \psi_1(\br,t) = 
 \frac{\dlt H[\eta,\psi_1]}{\dlt\psi^\dgr_1(\br,t)}  \; .
\ee

Let us introduce the normal density matrix
\be
\label{33}
\rho_1(\br,\br') \equiv \lgl \psi_1^\dgr(\br') 
\psi_1(\br) \rgl
\ee
and the so-called anomalous density matrix
\be
\label{34}
\sgm_1(\br,\br') \equiv \lgl \psi_1(\br') 
\psi_1(\br) \rgl \; .
\ee
Their diagonal elements define the density of uncondensed atoms and 
the anomalous average, respectively:
\be
\label{35}
\rho_1(\br) =  \lgl \psi_1^\dgr(\br) \psi_1(\br) \rgl \; ,
\qquad \sgm_1(\br) =  \lgl \psi_1(\br) \psi_1(\br) \rgl \; .
\ee
The local density of atoms is the sum
\be
\label{36}
\rho(\br) = \rho_0(\br) + \rho_1(\br) \;  .
\ee
We shall also need the triple correlator
\be
\label{37}
\xi(\br,\br') \equiv \lgl \psi_1^\dgr(\br') \psi_1(\br') 
 \psi_1(\br) \rgl \; .
\ee

With these notations, Eq. (31) results in the evolution equation 
for the condensate wave function
$$
i\; \frac{\prt}{\prt t} \; \eta(\br) = \left ( -\; 
\frac{\nabla^2}{2m} + U - \mu_0 \right ) \eta(\br) \; +
$$
\be
\label{38}
+ \; \int \Phi(\br-\br') \left [ \rho(\br') \eta(\br) +
\rho_1(\br,\br')\eta(\br') + \sgm_1(\br,\br')\eta^*(\br')
+ \xi(\br,\br') \right ] \; d\br' \; ,
\ee
while Eq. (32) yields the equation of motion for the operator 
of uncondensed atoms  
$$
i\; \frac{\prt}{\prt t} \; \psi_1(\br) = \left ( -\; 
\frac{\nabla^2}{2m} + U - \mu_1 \right ) \psi_1(\br) \; +
$$
\be
\label{39}
+ \; \int \Phi(\br-\br') \left [ | \eta(\br') |^2 
\psi_1(\br) + \eta^*(\br')\eta(\br) \psi_1(\br') + 
\eta(\br') \eta(\br) \psi_1^\dgr(\br') + 
\hat X(\br,\br') \right ] \; d\br' \; ,
\ee
in which
$$
\hat X(\br,\br') \equiv  
\psi_1^\dgr(\br') \psi_1(\br')\eta(\br) +
\psi_1^\dgr(\br') \eta(\br') \psi_1(\br) + 
\eta^*(\br') \psi_1(\br') \psi_1(\br) + 
\psi_1^\dgr(\br') \psi_1(\br') \psi_1(\br) \;  .
$$

For an equilibrium system, one has
$$
 \frac{\prt}{\prt t}\; \eta(\br) = 0 \qquad (equilibrium) \; .
$$
Then Eq. (38) reduces to the eigenvalue problem
$$
\mu_0 \eta(\br) = \left ( -\; \frac{\nabla^2}{2m} + U 
\right ) \eta(\br) \; +
$$
\be
\label{40}
+  \; \int \Phi(\br-\br') \left [ \rho(\br')\eta(\br) +
\rho_1(\br,\br')\eta(\br') + \sgm_1(\br,\br') \eta^*(\br') +
\xi(\br,\br') \right ] \; d\br' \; ,
\ee
defining the condensate wave function and the Lagrange multiplier 
$\mu_0$. 

Note that the Lagrange multipliers $\mu_0$ and $\mu_1$ do not 
need to coincide with each other. Their relation to the system 
chemical potential $\mu$ is given [13-15] by the equality
$$
\mu =\mu_0 n_0 + \mu_1 n_1 \;,
$$
in which the atomic fractions of condensed and uncondensed atoms, 
respectively, are
$$   
n_0 \equiv \frac{N_0}{N} \; , \qquad  
n_1 \equiv \frac{N_1}{N} \; .
$$
The formalism of this section provides the basis for the 
self-consistent theory of arbitrary nonuniform Bose-condensed 
systems [8,13-15].

\section{Hartree-Fock-Bogolubov approximation}

To proceed further, let us resort to the Hartree-Fock-Bogolubov 
(HFB) approximation, following the way of Refs. [13-15]. Then 
Hamiltonian (26) in the HFB approximation becomes 
$$
H_{HFB} = E_{HFB} + \int \psi_1^\dgr(\br) \left ( - \; 
\frac{\nabla^2}{2m} + U - \mu_1 \right ) \psi_1(\br) \; d\br \; +
$$
$$
+ \; \int \Phi(\br-\br') \left [ 
\rho(\br') \psi_1^\dgr(\br)\psi_1(\br) +
\rho(\br',\br) \psi_1^\dgr(\br' ) \psi_1(\br) +
\frac{1}{2} \; \sgm(\br,\br') \psi_1^\dgr(\br' ) \psi_1^\dgr(\br) +
\right.
$$
\be
\label{41}
\left. +  \frac{1}{2} \; 
\sgm^*(\br,\br') \psi_1(\br' ) \psi_1(\br) \right ] \; 
d\br d\br' \; ,
\ee
where
\be
\label{42}
E_{HBF} = H^{(0)} - \; \frac{1}{2} \int  \Phi(\br-\br') \left [
\rho_1(\br) \rho_1(\br') + |\rho_1(\br,\br')|^2 +
|\sgm_1(\br,\br')|^2 \right ] \; d\br d\br'\; ,
\ee
and the notation is introduced for the total normal density matrix
\be
\label{43}
\rho(\br,\br') \equiv \eta(\br) \eta^*(\br') + 
\rho_1(\br,\br') \; ,
\ee
and for the total anomalous density matrix
\be
\label{44}
\sgm(\br,\br') \equiv \eta(\br) \eta(\br') + 
\sgm_1(\br,\br') \;   .
\ee   
In the HFB approximation, the condensate-function equation (38) 
takes the form
$$
i\; \frac{\prt}{\prt t} \; \eta(\br) = \left ( - \; 
\frac{\nabla^2}{2m} + U - \mu_0 \right ) \eta(\br) \; +
$$
\be
\label{45}
+ \; \int \Phi(\br-\br') \left [
\rho(\br') \eta(\br) + \rho_1(\br,\br') \eta(\br') +
\sgm_1(\br,\br') \eta^*(\br') \right ] \, d\br ' \; .
\ee
And Eq. (39) for the field operator of uncondensed atoms reduces 
to
$$
i\; \frac{\prt}{\prt t} \; \psi_1(\br) = \left ( - \; 
\frac{\nabla^2}{2m} + U - \mu_1 \right ) \psi_1(\br) \; +
$$
\be
\label{46}
+ \; \int \Phi(\br-\br') \left [
\rho(\br') \psi_1(\br) + \rho(\br,\br') \psi_1(\br') +
\sgm(\br,\br') \psi_1^\dgr(\br') \right ] \; d\br ' \;  .
\ee

For an equilibrium system, the eigenvalue problem (40) becomes
$$
\mu_0 \eta(\br) = \left ( - \; \frac{\nabla^2}{2m} + U  
\right ) \eta(\br) \; +
$$
\be
\label{47}
+ \; \int \Phi(\br-\br') \left [
\rho(\br') \eta(\br) + \rho_1(\br,\br')\eta(\br') +
\sgm_1(\br,\br')\eta^*(\br') \right ]\; d\br ' \;  .
\ee

The HFB Hamiltonian (41) can be diagonalized by means of the general 
Bogolubov canonical transformations [16] that, in the used notation, 
read as 
$$
\psi_1(\br) = \sum_k \left [ b_k u_k(\br) + 
b_k^\dgr v_k^*(\br) \right ] \; ,
$$
\be
\label{48}
b_k = \int \left [ u_k^*(\br) \psi_1(\br) -
v_k^*(\br) \psi_1^\dgr(\br) \right ] \; d\br' \; ,
\ee
where $k$ is a multi-index. Since $\psi_1$ is the Bose operator, one 
has
$$
\sum_k \left [ u_k(\br) u_k^*(\br') -
v_k^*(\br) v_k(\br') \right ] = \dlt(\br-\br') \; ,
$$
\be
\label{49}
\sum_k \left [ u_k(\br) v_k^*(\br') -
v_k^*(\br) u_k(\br') \right ] = 0 \; .
\ee
And requiring that $b_k$ be also a Bose operator gives
$$
\int \left [ u_k^*(\br) u_p(\br) -
v_k^*(\br) v_p(\br) \right ]\; d\br = \dlt_{kp} \; ,
$$
\be
\label{50}
 \int \left [ u_k(\br) v_p(\br) -
v_k(\br) u_p(\br) \right ]\; d\br = 0\; .
\ee

The condition that Hamiltonian (41) be diagonalized by the Bogolubov 
transformation (48) is equivalent to the Bogolubov equations
$$
\int \left [ \om(\br,\br') u_k(\br') + \Dlt(\br,\br') v_k(\br') 
\right ] \; d\br' = \ep_k u_k(\br) \; ,
$$
\be
\label{51}
\int \left [ \om^*(\br,\br') v_k(\br') + \Dlt^*(\br,\br') u_k(\br') 
\right ] \; d\br' = -\ep_k v_k(\br) \; ,
\ee 
in which
$$
\om(\br,\br') \equiv \left [ - \; \frac{\nabla^2}{2m} + U(\br) -
\mu_1 + \int \Phi(\br-\br')\rho(\br')\; d\br' 
\right ] \; \dlt(\br-\br') \; +
$$
\be
\label{52}
  + \; \Phi(\br-\br') \rho(\br,\br')
\ee
and
\be
\label{53}
\Dlt(\br,\br') \equiv \Phi(\br-\br') \sgm(\br,\br') \; .
\ee
The Bogolubov equations (51) is the eigenproblem defining the 
Bogolubov functions $u_k({\bf r})$ and $v_k({\bf r})$ and the 
spectrum of collective excitations $\ep_k$.

As a result of the diagonalization, Hamiltonian (41) reduces to 
the Bogolubov form
\be
\label{54}
H_B = E_B + \sum_k \ep_k b_k^\dgr b_k \;  ,
\ee
where
\be
\label{55}
E_B \equiv E_{HFB} - \sum_k \ep_k \int 
| v_k(\br)|^2 \; d\br \; .   
\ee 
With the diagonal Hamiltonian (54), it is straightforward to 
calculate the required averages. Thus, for the distribution of 
excitations, one gets
\be
\label{56}
\pi_k \equiv \lgl b_k^\dgr b_k \rgl = \left ( e^{\bt\ep_k} 
- 1 \right )^{-1} .
\ee
The normal density matrix (33) is
\be
\label{57}
\rho_1(\br,\br') =\sum_k \left [
\pi_k u_k(\br) u_k^*(\br') +
(1 + \pi_k) v_k^*(\br) v_k(\br') \right ] \; .
\ee
And the anomalous density matrix (34) becomes
\be
\label{58}
\sgm_1(\br,\br') =\sum_k \left [
\pi_k u_k(\br) v_k^*(\br') +
(1 + \pi_k) v_k^*(\br) u_k(\br') \right ] \;  .
\ee
For the diagonal elements of Eqs. (57) and (58), we obtain
\be
\label{59}
\rho_1(\br) = \sum_k \left [ \pi_k | u_k(\br) |^2 + (1 +\pi_k)
| v_k(\br) |^2 \right ]
\ee
and, respectively,
\be
\label{60}
\sgm_1(\br) = \sum_k  (1 + 2\pi_k) u_k(\br) v_k^*(\br) \; .
\ee

\section{Local-density approximation}
 
When the spatial variation of the external nonuniform potential 
is sufficiently slow and the trap is sufficiently large, one can 
employ the local-density approximation [1,2,17], which is also 
called the semi-classical approximation and is widely used for 
describing trapped atoms [9,18,19].

In this approximation, one looks for the Bogolubov functions 
represented as the products
\be
\label{61}
u_k(\br) \equiv u(\bk,\br) \vp_k(\br) \; , \qquad
v_k(\br) \equiv v(\bk,\br) \vp_k(\br)
\ee
factorized with the plane waves
\be
\label{62}
\vp_k(\br) \equiv \frac{1}{\sqrt{V}}\; e^{i\bk\cdot\br} \; .
\ee
The amplitudes $u({\bf k},{\bf r})$ and $v({\bf k},{\bf r})$ are 
assumed to be slowly varying in space, as compared to the spatial  
variation of the plane wave,
\be
\label{63}
\frac{|\vec\nabla u(\bk,\br)|}{|u(\bk,\br)|} \; \ll \;
\frac{|\vec\nabla \vp_k(\br)|}{|\vp_k(\br)|} \; .
\ee

In the Bogolubov equations (51), one makes the replacements
$$
\int \om(\br,\br') u_k(\br') \; d\br ' \cong 
\om(\bk,\br) u_k(\br) \; ,
$$
\be
\label{64}
\int \om(\br,\br') v_k(\br') \; d\br ' \cong 
\om(\bk,\br) v_k(\br) \; ,
\ee
and 
$$
\int \Dlt(\br,\br') u_k(\br') \; d\br ' \cong 
\Dlt(\br) u_k(\br) \; ,
$$
\be
\label{65}
\int \Dlt(\br,\br') v_k(\br') \; d\br ' \cong 
\Dlt(\br) v_k(\br) \; ,
\ee
where the form
\be
\label{66}
\om(\bk,\br) \equiv \frac{k^2}{2m} + U(\br) + 2 \Phi_0 \rho(\br)
- \mu_1(\br)
\ee
is used, instead of Eq. (52), and the quantity 
\be
\label{67}
\Dlt(\br) \equiv [ \rho_0(\br) +\sgm_1(\br) ] \Phi_0
\ee
is used instead of Eq. (53), with the notation
\be
\label{68}
\Phi_0 \equiv \int \Phi(\br) \; d\br \; .
\ee
In what follows, we assume that $\Phi_0>0$.
Then the Bogolubov equations (51) reduce to the eigenproblem
$$
\left [ \om(\bk,\br) - \ep(\bk,\br) \right ] u(\bk,\br) +
\Dlt(\br) v(\bk,\br) = 0 \; ,
$$
\be
\label{69}
\Dlt^*(\br) u(\bk,\br) +
\left [ \om^*(\bk,\br) + \ep(\bk,\br) \right ] v(\bk,\br)
 = 0 \; .
\ee
The amplitudes $u({\bf k},{\bf r})$ and $v({\bf k},{\bf r})$ can 
be taken to be real and, in view of Eqs. (49), satisfying the 
relation
$$
u^2(\bk,\br) - v^2(\bk,\br) = 1 \; .       
$$

Solving eigenproblem (69) yields the Bogolubov spectrum of 
collective excitations
\be
\label{70}
\ep(\bk,\br) = \sqrt{\om^2(\bk,\br) - \Dlt^2(\bk,\br) } \; .
\ee
And for the amplitudes, we find
$$
u^2(\bk,\br) = 
\frac{\om(\bk,\br) + \ep(\bk,\br)}{2\ep(\bk,\br)} \; , 
\qquad
v^2(\bk,\br) = 
\frac{\om(\bk,\br)-\ep(\bk,\br)}{2\ep(\bk,\br)} \; ,
$$
\be
\label{71}
 u(\bk,\br) v(\bk,\br) = - \; 
\frac{\Dlt(\br)}{2\ep(\bk,\br)} \;  .
\ee

The necessary condition for the condensate existence [5], equivalent 
to the Hugenholtz-Pines theorem [20], requires that
\be
\label{72}
\lim_{k\ra 0} \ep(\bk,\br) = 0 \; , \qquad \ep(\bk,\br) \geq 0\; .
\ee 
This gives the Lagrange multiplier
\be
\label{73}
\mu_1(\br) = U(\br) + [ \rho_0(\br) + 2\rho_1(\br)
-\sgm_1(\br) ] \Phi_0 \; .
\ee 

Introducing the local sound velocity $c({\bf r})$ by the equation
\be
\label{74}
mc^2(\br) \equiv [ \rho_0(\br) + \sgm_1(\br) ] \Phi_0 
\ee
transforms Eq. (66) to
\be
\label{75}
\om(\bk,\br) = \frac{k^2}{2m} + mc^2(\br)  \; ,   
\ee 
while Eq. (67) becomes
\be
\label{76}
\Dlt(\br) = mc^2(\br) \; .
\ee
For the Bogolubov spectrum (70), we obtain the familiar expression
\be
\label{77}
\ep(\bk,\br) =
\sqrt{ c^2(\br) k^2 + \left ( \frac{k^2}{2m} \right )^2} \; ,
\ee
but with the local sound velocity.

Instead of distribution (56) for excitations, we have
\be
\label{78}
\pi(\bk,\br) = \{ \exp [\bt \ep(\bk,\br) ] -1 \}^{-1} \; ,
\ee
with the symmetry properties
\be
\label{79}
\pi(-\bk,\br) = \pi(\bk,\br)\; , \qquad
\ep(-\bk,\br) = \ep(\bk,\br) \; .
\ee
Using Eqs. (71), we find the normal density matrix (57) as
\be
\label{80}
\rho_1(\br,\br') = \sum_k n(\bk,\br) \vp_k(\br) \vp_k^*(\br') \; ,
\ee
with the distribution of atoms
\be
\label{81}
n(\bk,\br) = \frac{\om(\bk,\br)}{2\ep(\bk,\br)} \;
\coth \left [ \frac{\ep(\bk,\br)}{2T} \right ] \; - \;
\frac{1}{2}  \; .
\ee
For the anomalous density matrix (58), we get
\be
\label{82}
\sgm_1(\br,\br') = \sum_k 
\sgm(\bk,\br) \vp_k(\br) \vp_k^*(\br') \; ,
\ee
where
\be
\label{83}
\sgm(\bk,\br) = -\; \frac{mc^2(\br)}{2\ep(\bk,\br)} \;
\coth \left [ \frac{\ep(\bk,\br)}{2T} \right ] \; .
\ee
The diagonal elements of Eqs. (80) and (82) give the density of 
uncondensed atoms and the anomalous average, respectively:
\be
\label{84}
\rho_1(\br) = \frac{1}{V} \; \sum_k n(\bk,\br) \; , \qquad
\sgm_1(\br) = \frac{1}{V} \; \sum_k \sgm(\bk,\br)  \; .
\ee
The grand thermodynamic potential (9) becomes
\be
\label{85}
\Om = E_B + T \int \ln \left [ 1  -
\exp\{ -\bt \ep(\bk,\br) \} \right ] \;
\frac{d\bk}{(2\pi)^3} \; d\br \; .
\ee 
Employing the above formulas, one can calculate all thermodynamic 
characteristics.

\section{Particle fluctuations and stability}

Particle fluctuations are defined by the number-of-particle 
operator dispersion (11). The latter is proportional to the 
isothermal compressibility (12) which is to be thermodynamically 
normal for the stability of the system. 

In the grand canonical ensemble, used here, the compressibility 
is 
\be
\label{86}
\kappa_T = \frac{\Dlt^2(\hat N)}{\rho T N} \; , \qquad
\rho \equiv \frac{1}{V} \int \rho(\br) \; d\br \; .
\ee
In the canonical ensemble, the number of particles is fixed. 
However, it would be incorrect to conclude that the compressibility 
then is zero. Expression (86) is not defined for the canonical 
ensemble. Instead, one should use the formula
$$
 \kappa_T =  \frac{1}{V} 
\left ( \frac{\prt^2 F}{\prt V^2} \right )^{-1}_{TN} \; ,
$$
where $F$ is the canonical free energy.

In the same way, it would be principally wrong to state that, 
fixing the number of atoms $N$ and setting the particle 
dispersion (11) to zero, would result in the equality 
$\Dlt^2(\hat N_0)=\Dlt^2(\hat N_1)$  that would define the 
condensate fluctuations $\Dlt^2(\hat N_0)$ in the canonical 
ensemble by calculating the dispersion $\Dlt^2(\hat N_1)$ 
of uncondensed atoms. This way of reasoning is wrong because 
the particle dispersion (11) is not defined for the canonical 
ensemble. In addition, calculating $\Dlt^2(\hat N_1)$ in the 
Fock space has nothing to do with the canonical ensemble, 
as far as in the Fock space, the number of particles is not 
fixed. 

In the Fock space, the correct conclusion [10] for the particle 
fluctuations, after using the Bogolubov shift (17), is that the 
condensate fluctuations are zero,
\be
\label{87}
\Dlt^2(\hat N_0) = 0 \; ,
\ee
and that the particle fluctuations are completely due to those 
of uncondensed atoms,
\be
\label{88}
\Dlt^2(\hat N) = \Dlt^2(\hat N_1) \;  .
\ee
Expressions (87) and (88), for large particle numbers $N$ are 
asymptotically exact [10].

It is also important that the fluctuations of the total number of 
particles are thermodynamically normal if and only if the fluctuations 
of both the condensate fraction as well as of uncondensed atoms are 
thermodynamically normal. And, vice versa, the fluctuations of the 
total number of particles are thermodynamically anomalous if and only 
if a least one of the fractions produces thermodynamically anomalous 
fluctuations. The corresponding theorem has been rigorously proved 
in Refs. [6-8]. 

When the HFB approximation is involved, Eq. (88) is not convenient 
to use, since the HFB approximation is an effective second-order 
approximation with regard to the powers of the field operators of 
uncondensed atoms $\psi_1$. In that second-order approximation, the 
quantity $\hat {N}_1^2$ is not defined, being of the fourth order  
with respect to $\psi_1$. But we can employ another way of calculating 
the particle dispersion (11). We may notice that the latter can be 
expressed through the density-density correlation function
\be
\label{89}
D(\br,\br') \equiv \lgl \hat\psi^\dgr(\br) \hat\psi(\br) 
\hat\psi^\dgr(\br') \hat\psi(\br') \rgl
\ee
as
\be
\label{90}
\Dlt^2(\hat N) = \int \left [ D(\br,\br') - \rho(\br)\rho(\br') 
\right ] \; d\br d\br' \;  .
\ee  
The density-density correlation function (89) is related to the 
diagonal correlation function
\be
\label{91}
C(\br,\br') \equiv \lgl \hat\psi^\dgr(\br) \hat\psi^\dgr(\br')
\hat\psi(\br') \hat\psi(\br) \rgl
\ee
by the equality
\be
\label{92}
D(\br,\br') =\rho(\br) \dlt(\br-\br') + C(\br,\br') \; .
\ee 
In turn, the diagonal correlation function (91) is connected with 
the pair correlation function
\be
\label{93}
g(\br,\br') \equiv \frac{\lgl \hat\psi^\dgr(\br) \hat\psi^\dgr(\br')
\hat\psi(\br') \hat\psi(\br) \rgl}{\rho(\br)\rho(\br') }
\ee
through the relation
\be
\label{94}
C(\br,\br') = \rho(\br)\rho(\br') g(\br,\br') \;  .
\ee
Therefore the particle dispersion (90) can be represented as
$$
\Dlt^2(\hat N) = N + \int \rho(\br) \rho(\br')
\left [ g(\br,\br') -1 \right ] \; d\br d\br' =
$$
\be
\label{95}
= N + \int [ C(\br,\br') - \rho(\br) \rho(\br') ] \; 
d\br d\br' \; .
\ee
This formula is valid for arbitrary nonuniform systems.

Accomplishing the Bogolubov shift (17) in Eq. (91) yields
$$
C(\br,\br') = \rho(\br) \rho_0(\br') + \rho_0(\br)\rho_1(\br')
+ 
$$
$$
+ 2 {\rm Re} \left [ \eta^*(\br) \eta^*(\br') \rho_1(\br,\br')
+ \eta^*(\br) \eta^*(\br')\sgm_1(\br,\br') + 
\eta^*(\br) \xi(\br,\br') + \eta^*(\br) \xi(\br',\br) 
\right ] \; + 
$$
\be
\label{96}
+ \; C_1(\br,\br') \;  ,
\ee
where
\be
\label{97}
C_1(\br,\br') \equiv \lgl \psi_1^\dgr(\br) \psi_1^\dgr(\br')
\psi_1(\br') \psi_1(\br)  \rgl \; .
\ee 
In the HFB approximation, the triple correlator (37) is zero, 
while the correlation function (97) becomes
\be
\label{98}
C_1(\br,\br') = \rho_1(\br) \rho_1(\br') +
| \rho_1(\br,\br') |^2  + | \sgm_1(\br,\br') |^2 \;  .
\ee
Then the correlation function (96) reduces to
$$
C(\br,\br') = \rho(\br) \rho(\br') + 2 {\rm Re} \left [ 
\eta^*(\br) \eta(\br') \rho_1(\br,\br')
+ \eta^*(\br) \eta^*(\br')\sgm_1(\br,\br') \right ] +
$$
\be
\label{99}
+ | \rho_1(\br,\br') |^2  + | \sgm_1(\br,\br') |^2   \; .
\ee
As is stressed above, the HFB approximation is of second order with 
respect to the operators of uncondensed atoms. The terms, containing 
higher orders are not defined in this approximation and have to be 
omitted. This concerns the last two terms in Eq. (99). At the same 
time, in the frame of this approximation for an equilibrium system, 
the function $\eta({\bf r})$ in the second and third terms of Eq. 
(99) can be replaced by the real value $\sqrt{\rho({\bf r})}$. As 
a result, for the particle dispersion (95), we obtain
\be
\label{100}
\Dlt^2(\hat N) = N + 2 \int \sqrt{\rho(\br) \rho(\br') } \;
\left [ \rho_1(\br,\br') + \sgm_1(\br,\br') \right ] \;
d\br d\br' \; .
\ee
In the spirit of the local-density approximation, the latter 
expression can be rewritten as 
\be
\label{101}
\Dlt^2(\hat N) = N + 2 \int \rho(\br)  \;
\left [ \rho_1(\br,\br') + \sgm_1(\br,\br') \right ] \;
d\br d\br' \;  .
\ee
Invoking Eqs. (80) and (82) gives
\be
\label{102}
\int \rho_1(\br,\br') \; d\br' = \lim_{k\ra 0} n(\bk,\br) \; , 
\qquad  
\int \sgm_1(\br,\br') \; d\br' = \lim_{k\ra 0} \sgm(\bk,\br) \; .
\ee
Hence dispersion (101) reads as
\be
\label{103}
\Dlt^2(\hat N) = N + 2 \int \rho(\br) \lim_{k\ra 0}
\left [ n(\bk,\br) + \sgm(\bk,\br) \right ] \; d\br \;  .
\ee
Using Eqs. (81) and (83), we get
$$
\lim_{k\ra 0} \left [ n(\bk,\br) + \sgm(\bk,\br) 
\right ]  = \frac{1}{2} \left [ \frac{T}{mc^2(\br) } \; - \; 
1 \right ] \; .       
$$
It is worth emphasizing that both Eqs. (81), as well as (83), 
diverge in the long-wave limit $k \rightarrow 0$. But their 
divergences, being of opposite signs, cancel each other. This 
means that taking into account the anomalous average (83) is 
of principal importance. Without it, the dispersion (103) would 
diverge, and the compressibility (86) would be infinite, which 
implies the system instability.

In that way, the particle fluctuations are described by the 
dispersion
\be
\label{104}
\Dlt^2(\hat N) = \int \frac{T\rho(\br)}{mc^2(\br) } \; d\br \;  .
\ee
Therefore, the compressibility (86) is
\be
\label{105}
\kappa_T = \frac{1}{m\rho N} \int \frac{\rho(\br)}{c^2(\br) }
\; d\br  .
\ee
These formulas show that particle fluctuations are 
thermodynamically normal and compressibility (105) is finite, 
even in the thermodynamic limit. At the critical temperature 
$T_c$, where $c({\bf r})$ tends to zero, the compressibility 
diverges as $T \rightarrow T_c$. This divergence is typical 
for the point of a second-order phase transition, where the 
system is unstable. However everywhere below $T_c$, the 
compressibility is finite and fluctuations are thermodynamically 
normal. This conclusion does not depend on the statistical ensemble 
involved. Both canonical and grand canonical ensembles for Bose 
systems produce the results coinciding in the thermodynamic limit 
[21], provided all calculations are done  correctly, by using the 
representative statistical ensembles [8,13-15,22].

The compressibility (105) is a directly observable quantity, 
being related to the structure factor (14). The latter can be 
measured by studying light scattering from ultracold atomic gases 
[23,24]. No one scattering experiment with Bose-condensed systems 
has ever showed a thermodynamically anomalous structural factor.

Thermodynamically anomalous fluctuations for a stable statistical 
system can arise solely because of incorrect calculations. For 
example, if in the HFB or Bogolubov approximation, one includes 
in calculations the fourth- or higher-order terms, which are not 
defined  in the second-order approximation, then one can get any 
kind of thermodynamic anomalies. However, such anomalies are 
physically meaningless, being mathematically wrong. For correctly 
calculating  the fourth-order terms, one has to use a fourth-order 
Hamiltonian. 

Nonperturbative thermodynamics of an interacting Bose gas, for all 
temperatures below $T_c$, has recently been studied by Floerchinger 
and Wetterich [25] by using renormalization-group techniques, which 
effectively take into account all orders of field operators. Their 
results confirm that the compressibility is finite everywhere below 
$T_c$, hence, all particle fluctuations are thermodynamically normal.   

{\it Acknowledgement}. Financial support from the Russian 
Foundation for Basic Research is appreciated.

\end{document}